\documentclass[prb,showpacs,floatfix,twocolumn]{revtex4}
\usepackage{amsfonts}
\usepackage{stmaryrd}
\usepackage{bbm}
\usepackage{mathrsfs}
\usepackage{tipa}
\usepackage{amssymb}
\usepackage{txfonts}
\usepackage{graphicx}
\usepackage{dcolumn}
\begin{document}

\newcommand*{\cm}{cm$^{-1}$\,}


\title{Optical study of the multiple charge density wave transitions in ErTe$_3$}
\author{B. F. Hu}
\author{B. Cheng}
\author{R. H. Yuan}
\author{T. Dong}
\author{A. F. Fang}
\author{W. T. Guo}
\author{Z. G. Chen}
\author{P. Zheng}
\author{Y. G. Shi}
\author{N. L. Wang}

\affiliation{Beijing National Laboratory for Condensed Matter
Physics, Institute of Physics, Chinese Academy of Sciences,
Beijing 100190, People's Republic of China}
%


\begin{abstract}
We present an optical spectroscopy study on singe crystalline
ErTe$_3$, a rare-earth-element tri-telluride which experiences two
successive charge density wave (CDW) transitions at T$_{c1}$ = 267 K
and T$_{c2}$ = 150 K. Two corresponding gap features, centered at
2770 \cm ($\sim$ 343 meV) and 890 \cm ($\sim$ 110 meV) respectively,
are clearly seen in ordered state. A pronounced Drude component,
which exists at all measurement temperatures, demonstrates the
partial gap character of both CDW orders. About half of the
unmodulated Fermi surface (FS) remains in the CDW state at the
lowest measurement temperature. The study also reveals that
fluctuation effect is still prominent in this two-dimensional (2D)
material.
\end{abstract}


\maketitle

\section{Introduction}
Charge density waves (CDWs), as typical broken symmetry states in
solids,\cite{densitywave} have received considerable attention in
the past decades. To date, the CDW orders are accepted to be
predominantly driven by the nesting of the metallic Fermi surfaces
(FSs). Such nesting would cause a divergence of the electronic
susceptibility and response function at the nesting wave vector
\textbf{k}$_F$, resulting in instability of the FS. An energy gap
in the single particle excitation spectrum would open near the
Fermi level, leading to lowering of the total energy of the
system. In real space, the instability would result in a spatially
periodical modulation of charge density distribution
$\rho(\textbf{r})$ with a period of $\lambda$ =
$\pi/\textbf{k}_F$. Mediated by the electron-phonon coupling, it
also causes a displacement or superstructure distortion of the
underlying lattice. Up to now, a great deal of experimental and
theoretical investigations have been done aiming at exploring and
understanding the intriguing collective quantum phenomena in a
variety of CDW materials. The low dimensional systems, which have
a highly anisotropic crystal structures, are especially concerned
due to the relatively easy realization of the FS nesting
condition. Nevertheless, except for the one-dimensional electron
system, perfect nestings among the entire FS could hardly be
achieved in real materials. More often, only parts of the FS are
nested, which could be connected by a wave vector that matches
with a peculiar phonon anomaly. In this circumstance, only partial
energy gap opens on the FS in the optimally nested area and some
metallic channels are saved in ordered state.

The discovery of the CDW character in the rare-earth tri-telluride
RTe$_3$ (R=Y, La, and rare earth elements) systems \cite{TEMCDW} has
provided an ideal opportunity to probe the CDW properties, which
have been well characterized by the FS nesting driven mechanism.
\cite{theory,SmTe3,ARPESCeTe3400meV,ARPESCeTe3,ARPESRTe3,ARPES2CDW}
RTe$_3$ has a layered, weakly orthorhombic crystal structure (No. 63
space group \emph{Cmcm}), \cite{structure,structure2} making up of
two nearly square Te nets, which are separated by an insulating
corrugated RTe slab. The former undertakes the metallic conductive
task while the latter plays a role of charge reservoirs.
\cite{structure2,ARPESCeTe3400meV} The lattice of the compounds have
an incommensurate modulation with a wave vector of approximately
2/7\emph{c}$^{\ast}$ in the CDW state. \cite{TEMCDW} The transition
temperature decreases monotonically on traversing the lanthanides
series from lighter rare-earth to heavier ones,
\cite{material,quenchCDW,2CDW} which is slightly below the room
temperature for the heavy rare-earth RTe$_3$ \cite{2CDW} but
believed to be even above the melting point for the first several
light rare-earth compounds.\cite{SmTe3} ARPES study has well
demonstrated the evolution of the CDW properties, \cite{ARPESRTe3}
including the effect of chemical pressure on the energy gaps and the
distribution of the order parameter in reciprocal space. Meanwhile,
in consistent with the rather high transition temperatures, the
energy gap scale has been reported to have an order of several
hundred meV.
\cite{SmTe3,ARPESCeTe3400meV,ARPESCeTe3,RTe3che.pres,ARPESRTe3,opticalCeTe3}

Recently, apart from the above mentioned higher transitions, another
anomaly was observed in the transport measurement \cite{2CDW} of the
heavy rare-earth RTe$_3$ (R = Dy, Ho, Er, Tm) materials, which
reveals that there exists another transition at a lower temperature.
Then the notion is further confirmed by the following X-ray
diffraction study. \cite{2CDW} A group of superlattice peaks, which
develop below the transition, indicate that it has a CDW origin. The
study reflects that a lattice modulation, which holds a wave vector
along \emph{a}$^{\ast}$ axis, occurs perpendicular to the first one.
Subsequent ARPES investigation of ErTe$_3$, \cite{ARPES2CDW} one of
the heavy rare-earth tri-telluride which has two well separated
transitions at 267 K and 155 K, observed multiple gap features at 10
K, which strengthened the notion that multiple CDW orders coexist in
the system. The unmodulated spectral weight, which locates in the
neighborhood of \emph{c}$^{\ast}$ axis in reciprocal space, was
largely removed from the FS due to the first CDW gap formation,
which is consistent with the previous ARPES studies on light
rare-earth tri-telluride. Furthermore, the intensity on the original
FS in the vicinity of \emph{a}$^{\ast}$ axis is also clearly
reduced, and then it is ascribed to the development of the second
CDW order. The two energy gaps revealed by ARPES measurement are
$\Delta_1$ = 175 meV and $\Delta_2$ = 50 meV, respectively.
\cite{ARPES2CDW}

Optical spectroscopy is a powerful bulk sensitive technique to
detect the energy gaps in ordered state and yields a great wealth of
information in CDW systems. The recent report of optical study on
ErTe$_3$ \cite{opticsHoEr} suggests a partial gap formation roughly
at 3000 \cm at low temperatures. Although the compound experiences
two well separated CDW transitions, \cite{2CDW} which are both
believed to locate at rather high energy scales, \cite{ARPES2CDW} no
information responsible for the second CDW transition was found in
earlier optical measurement. Here we present an optical study on
single crystalline ErTe$_3$. In sharp contrast to the previous
assertion, the temperature dependent spectra clearly reflect the
development and evolution of both CDW orders. The first CDW
transition has an energy gap 2$\Delta_1$ = 2770 \cm ($\sim$ 343 meV)
and the second one 2$\Delta_2$ = 890 \cm ($\sim$ 110 meV) in the CDW
state, both being in good agreement with the ARPES measurement.

\section{\label{sec:level2}EXPERIMENT AND RESULTS}
The as-grown single crystalline samples of ErTe3, on which all the
present measurements performed, were grown by a self-flux method
which was illustrated in detail in our earlier work.
\cite{opticalCeTe3} The plate-like single crystal with shining
golden surface (about 4$\times$4 mm$^2$) is flexible and much
convenient to cleave. We noticed that the sample is much more air-
and moisture-sensitive than the light rare-earth element RTe$_3$
compounds. \cite{opticalCeTe3} After an exposure in air just a few
days, the samples became soft and frail associated with a
progressively darker surface. In order to avoid the deterioration,
we have tried our best to diminish the samples' exposure time in
air. Once the growth process came to the end and the single crystal
was obtained, we carried out the measurements as soon as possible.
Before each measurement process, the newly cleaved fresh surface is
obtained. Meanwhile, each measurement was carried out several times
to guarantee the repeatability and identity of the experimental
results.

The temperature dependence of the in-plane (ac-plane) dc
conductivity is obtained by the standard four-probe method and
plotted in figure 1. In consistent with previous studies,
\cite{2CDW} two obvious anomalies near T$_{c1}$ = 267 K and T$_{c2}$
= 150 K exhibit in the curve which are ascribed to the two CDW
transitions. No clear hysteresis is observed between cooling and
warming process which is a clue that both transitions are
second-order like. \cite{2CDW,ARPES2CDW} The overall metallic
behavior reveals that the FS is only partially affected across the
two CDW transitions. \cite{ARPES2CDW}

\begin{figure}[b]
\includegraphics[width=7.8cm,clip]{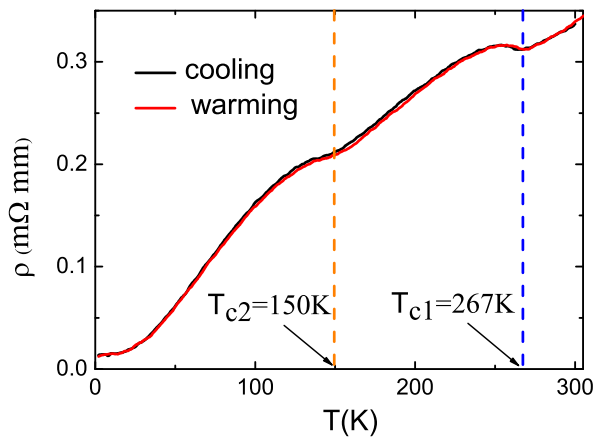}
\caption{(Color online) The temperature dependent in-plane
(ac-plane) resistivity of single crystal ErTe$_3$ for both cooling
and warming process. Two transitions emerge at T$_{c1}$ = 267 K
and T$_{c2}$ = 150 K respectively.}
\end{figure}

The optical reflectivity measurement was bought out on a combination
of Bruker IFS 113v and 80v/s spectrometers in a frequency range from
40 to 40000 cm$^{-1}$. An \textit{in situ} gold and aluminium
overcoating technique was used to get the reflectivity R($\omega$).
Kramers-Kronig transformation of R($\omega$) is employed to get the
real part of the conductivity spectra $\sigma_1(\omega)$. A
Hagen-Rubens relation was used for the low frequency extrapolation.
In accordance with our earlier work, \cite{opticalCeTe3} a constant
value of high frequency extrapolation was used up to 100000 \cm,
above which an $\omega^{-4}$ relation was employed.

The main panel of figure 2 focuses on the low frequency range of the
optical reflectivity spectra R($\omega$) up to 5000 \cm, while its
inset displays the expanded range up to 36000 \cm at two selected
temperatures. The most significant character in R($\omega$) spectra
is the development of a strong dip feature located in the
mid-infrared region upon cooling. We notice that there already
exists a board suppression near 1500 \cm in the room temperature
spectrum. During cooling precess, the dip feature, which has been
rather obvious at 200 K, becomes more and more dramatic. At the
lowest temperature it is rather pronounced. Furthermore, the center
of the depletion shows a notable shift towards higher energies,
roughly at 2500 \cm at 10 K. Additionally, at a lower energy scale a
second apparent suppression near 900 \cm is observed in R($\omega$)
at 10 K. In opposition to the first one, it is much less pronounced
and could not be clearly identified at the measurement temperatures
above 100 K. It is worthy noting that the suppressions are typical
features of FS gap formation caused by density wave instabilities.
\cite{opticalCeTe3,opticalTaS2,opticalBaFe2As2,opticalNaFeAs} As we
shall elaborate below, these are the behaviors in reflectance
corresponding with the two CDW orders respectively.

\begin{figure}[t]
\includegraphics[width=7.8cm,clip]{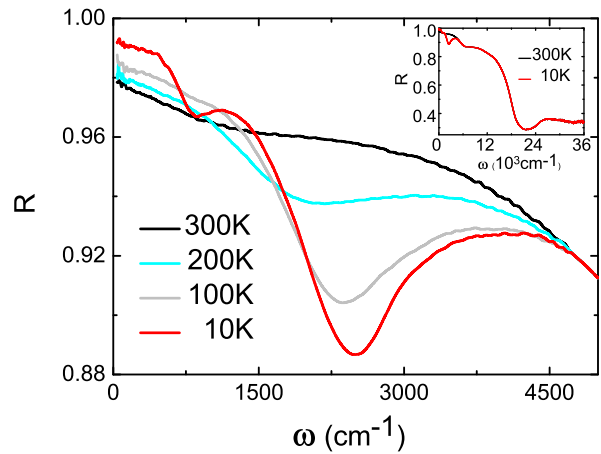}
\caption{(Color online) The reflectivity spectra R($\omega$) of
ErTe$_3$ at varied temperatures in the range from 40 \cm to 5000
\cm. Inset: the reflectivity spectra at two representative
temperatures in an expanded range up to 36000 \cm.}
\end{figure}

From the inset of figure 2, it could be concluded that all the
spectra at varied temperatures coincide together roughly at 4700
\cm, above which, the reflectance decreases slowly overall at first
and displays an obvious broad kink near 6700 \cm. Different from the
above mentioned two suppressions, the kink is already present at
room temperature and exhibits no temperature dependence. As
discussed in the next section, it should be ascribed to the
inter-band transition. As frequency continues to proceed, the
spectra drop sharply and form a well-defined reflectance edge
(plasma edge), then merge with the background reflectance
contributed from the inter-band transitions. The well-defined plasma
edge, together with the rather high reflectivity in the extremely
low frequency region, demonstrate that the compound is a good metal
even if it enters the CDW state. Below the two CDW energy scales,
the reflectance increases monotonically when lowing temperatures,
which is consistent with the dc resistivity measurement. It
reinforces the metallic character of the material down to low
temperatures, thus the FS is only partially gapped associated with
the development of the CDW orders.

\begin{figure}[t]
\includegraphics[width=7.8cm,clip]{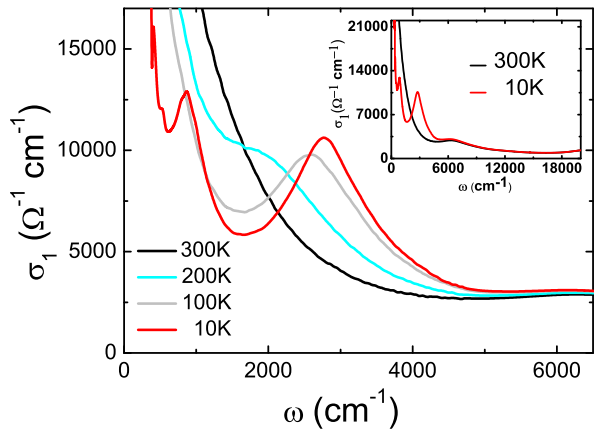}
\caption{(Color online) The temperature dependence of the real
part of the conductivity $\sigma_1(\omega)$ up to 6500 \cm. The
inset shows the spectra at 10 K and 300 K over a broad frequency
range.}
\end{figure}

The real part of conductivity $\sigma_1(\omega)$ is displayed in
figure 3 with its inset providing a global description in full
range. A pronounced Drude component exists at all temperatures,
confirming the metallic behaviors. While in the CDW state, a notable
mid-infrared peak near 2800 \cm emerges in $\sigma_1(\omega)$ at 10
K. As temperature increases, the peak shifts slightly to lower
energies and decreases in magnitude, tends to vanish at room
temperature spectra eventually. Furthermore, another similar peak
feature at a lower energy scale appears only at 10 K. Corresponding
with the two dip characters in R($\omega$), both peak features in
$\sigma_1(\omega)$ indicate the CDW gap properties. Additionally, we
note that the inter-band transition peak near 6700 \cm is also
clearly seen at all temperatures.

\section{\label{sec:level2}DISCUSSION}

\begin{table*}[htbp]
\begin{center}
\newsavebox{\tablebox}
\begin{lrbox}{\tablebox}
\begin{tabular}{*{14}{m{8mm}}}
\hline \hline\
{}&$\omega_{p}$&$\gamma_{D}$&$\omega_1$&$\gamma_1$&$S_1$&$\omega_2$&$\gamma_2$&$S_2$&$\omega_3$&$\gamma_3$&$S_3$\\[4pt]
\hline
300K&47&1.00&--&--&--&--&--&--&6.7&5.0&24\\[4pt]
200K&41&0.55&--&--&--&2.0&2.1&28&6.7&4.8&23\\[4pt]
100K&39&0.41&--&--&--&2.6&2.1&31&6.7&4.7&23\\[4pt]
10K&32&0.15&0.89&0.65&19&2.8&1.7&31&6.7&4.6&23\\[4pt]
\hline \hline
\end{tabular}
\end{lrbox}
\caption{Temperature dependence of the plasma frequency $\omega_p$
and scattering rate $\gamma_D$=1/$\tau_D$ of the Drude term, the
resonance frequency $\omega_i$, the width $\gamma_i$=1/$\tau_i$
and the square root of the oscillator strength $S_i$ of the
Lorentz component(all entries in 10$^3$ \cm). One Drude mode is
employed for all four measurement temperatures. Two Lorentz terms
responsible for the CDW orders are added at low temperatures. The
lowest energy inter-band transition is also displayed.}
\scalebox{1.0}{\usebox{\tablebox}}
\end{center}
\end{table*}

To analyze the evolution of both the itinerant carrier resonance and
the CDW gap excitations in a  quantitative way, the
$\sigma_1(\omega)$ spectra are fitted by the Drude-Lorentz model in
the whole frequency range for all four temperatures:

\begin{equation}
\epsilon(\omega)=\epsilon_\infty-{{\omega_p^2}\over{\omega^2+i\omega/\tau_D}}+\sum_{i=1}^N{{S_i^2}\over{\omega_i^2-\omega^2-i\omega/\tau_i}}.
\label{chik}
\end{equation}

Here, $\epsilon_\infty$ is the dielectric constant at high energy,
the middle and last terms are the Drude and Lorentz components,
respectively. A Drude term is employed to reproduce the metallic
response and a selection of Lorentz components are used for the
finite frequency excitations, including the double peak gap features
below the transition temperatures, an evident near-infrared peak
centered roughly at 6700 \cm, and two high energy ingredients above
25000 \cm for inter-band transitions. The fit results at 10 K are
presented in the main panel of figure 4 with its inset revealing the
room temperature counterpart. In addition, the temperature
dependence of the fit parameters, apart from the two high energy
constant inter-band transitions, are listed in table 1. In sharp
contrast to the other terms of which the parameters experience
remarkable changes at varied temperatures, the parameters of the
third Lorentz component almost exhibit no temperature dependence.
The central peak position is roughly pinned at 6700 \cm ($\sim$ 830
meV). The charge excitation locates at such a high energy scale that
it most likely originates from the inter-band transitions. In the
following part the discussions will be focused on the plasma
frequencies, Drude scattering rates and the two CDW gaps.

\begin{figure}[b]
\includegraphics[width=7.8cm,clip]{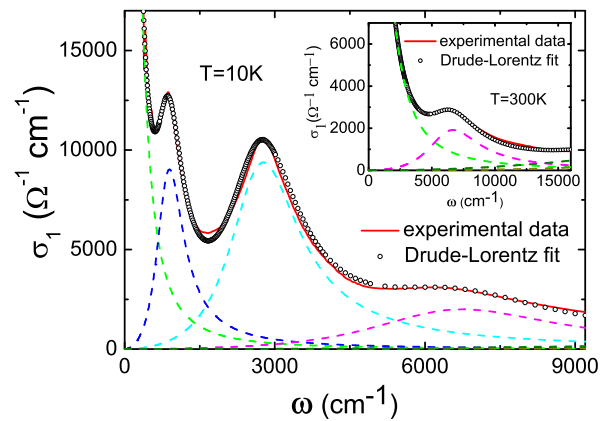}
\caption{(Color online) The experimental data of
$\sigma_1(\omega)$ at 10 K and the Drude-Lorentz fit results. Each
component of the fit displays at the bottom. Inset shows the
corresponding results at 300 K.}
\end{figure}

It is well known that the free carrier resonance governs the Drude
component behaviors. The Drude spectral weight determines the plasma
frequency $\omega_p$, the square of which is proportional to the
effective carrier density \emph{n}/\emph{m}$^{\ast}$ (where
\emph{m}$^{\ast}$ is the effective carrier mass). In addition, the
width of the Drude peak is just the scattering rate $\gamma$ =
1/$\tau$ (where $\tau$ is the average life time of free carriers).
The plasma frequency $\omega_p$ = 47100 \cm at 300 K reduces to
32000 \cm at 10 K. Provided that the effective mass of free carriers
remains the same at different temperatures, the area of the FS at 10
K will be 46\% of that at room temperature, meaning that about half
of the FS is gapped away from E$_F$. For the sake of verification
and comparison, another approach is adopted to estimate the ungapped
fraction of the FS: \cite{opticsHoEr}

\begin{equation}
\Phi={{\omega_p^2}\over{\omega_p^2+\sum_{i=1}^2{S_i^2}}}.
\label{chik}
\end{equation}

Where $\omega_p$ is the plasma frequency, $S_i$ is the mode strength
of the \emph{i}th Lorentz peak. The calculation gives a value $\Phi$
= 44$\%$ at 10 K. The above conclusion is well evidenced by the
nearly same value.

Unlike the relatively moderate change of the plasma frequency, the
scattering rate experiences much stronger variations. The room
temperature value $\gamma$ = 1025 \cm drops to 150 \cm at 10 K,
namely which reduces by about 85$\%$. It seems that the openings of
partial CDW gaps remove the electrons near Fermi level which undergo
stronger scattering. \cite{opticalCeTe3} Associated with the loss of
scattering channels, the metallic behavior reveals an enhancement of
the dc conductivity down to low temperatures.

Now the emphasis is assigned to the discussion of the CDW
properties. At the lowest temperature, $\sigma_1(\omega)$ spectra
develop two typical gap features, the first one centered at 2770
\cm ($\sim$ 343 meV) and the second 890 \cm ($\sim$ 110 meV). It
should be mentioned that the electrodynamics of density wave
ground states have been well established and understood. Due to
the first case coherence factor in the electrodynamic response,
optical conductivity $\sigma_1(\omega)$ in density wave state
shows a sharp maximum at the gap frequency.
\cite{opticalBaFe2As2,Degiorgi} At finite temperatures, the
thermally excited electrons contribute a low frequency tail.
Therefore, the double peak features can be identified as the
corresponding CDW behaviors in optical spectroscopy. Meanwhile,
the peak positions indicate the two CDW gaps. Combined with the
transition temperatures, we can get the ratio 2$\Delta$/
\emph{k}$_ B$T$_{CDW}$ = 14.9 for the first CDW order and 8.5 for
the second one. Both ratios are much higher than the weak coupling
theory predictions. It is worthy pointing out that both CDW gap
values obtained from the optical spectroscopy measurement match
well with that extracted from ARPES experiments, \cite{ARPES2CDW}
where a larger gap $\Delta_1$ = 175 meV and a smaller one
$\Delta_2$ = 50 meV are obtained. Double of both values are almost
identical with our results. Thus the present optical study is in
good agreement with the ARPES measurement. In the previous optical
report on the same compound, \cite{opticsHoEr} only one gap
feature near 3000 \cm was found. As we have mentioned above, the
material is air- and moisture-sensitive. The surface degradation,
if exposing in air for a bit longer time, may cause the
discrepancy.

\begin{figure}[t]
\includegraphics[width=7.8cm,clip]{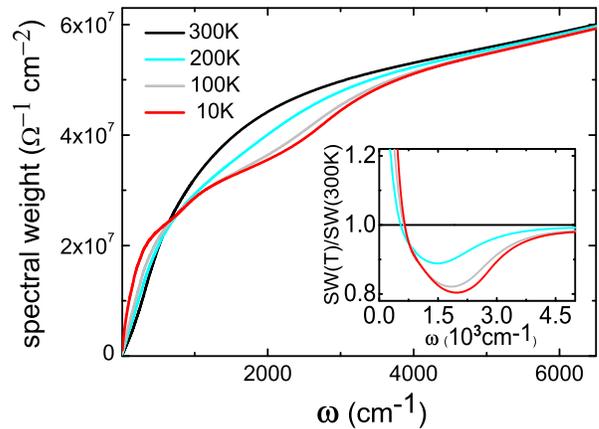}
\caption{(Color online) The temperature dependence of the spectral
weight up to 6500 \cm. Inset: the normalized spectral weight
SW(T)/SW(300K) up to 5000 \cm.}
\end{figure}

The evolution of the gap properties is apparent at varied
temperatures. During warming process, in $\sigma_1(\omega)$ spectra
the larger gap peak exhibits a sizable shift towards lower energies
and decreases in magnitude, becomes invisible at room temperature at
last. On the other hand, the corresponding suppression in
R($\omega$), as mentioned above, is clearly present at 300 K. It is
most likely that the weak character due to Lorentz excitation in the
room temperature spectra is covered by the high frequency tail of
the large Drude component. \cite{opticsHoEr} They match so well that
we could not distinguish the charge excitation across the CDW gap
from the itinerant carrier response. However, the transition
temperature T$_{c1}$ = 267 K in dc conductivity is well below the
room temperature. The fact that CDW characters emerge above the
transition supports the notion in this two-dimensional compound
fluctuation effect is still prominent. Such effect is also confirmed
by the observation in X-ray diffraction experiment \cite{2CDW} of
superlattice peaks with broadened width well above T$_{c1}$. In
addition, although the second transition temperature T$_{c2}$ = 150
K is well above 100 K, the corresponding CDW gap excitation which
exists in $\sigma_1(\omega)$ at 10 K could not be identified at 100
K. In contrast to the pronounced peak feature of the first CDW gap,
the peak magnitude corresponding with the second CDW order is much
weaker. Meanwhile, the peak position centered at 890 \cm is very
close to the giant zero energy excitations. So we believe that the
disappearance of the second CDW peak in $\sigma_1(\omega)$ at 100 K
has the same reason with the invisibility of the first one at 300 K.

To gain more insight into the band structure variation accompanied
with the CDW gap formations, a spectral weight analysis is present.
The spectral weight as a function of frequency is plotted in figure
5 with the normalized one showing in the inset. In the extremely low
frequency range, in consistent with the overall metallic behavior
observed in dc resistivity measurement, the spectral weight
increases monotonically with decreasing temperatures, due to the
narrowing and rising of the Drude component. In the mid-infrared
frequency range roughly between 700 \cm and 4500 \cm, the spectral
weight at lower temperatures becomes smaller than that at higher
ones. The redistribution clearly indicates the CDW gap formation.
The inset of figure 5 shows that a notable minimum emerges in the
normalized spectral weight in this area. At 10 K a minimal value
82\% reaches at $\omega_{min}$$\approx$ 2000 \cm. Consisting with
the movement of the peak position in $\sigma_1(\omega)$,
$\omega_{min}$ shifts remarkably to low frequency with increasing
temperatures, indicating that the CDW gaps tend to close. Above
$\omega_{min}$ the spectral weight differences between different
temperatures diminish gradually, suggesting a spectral weight
transfer from the low frequency region below $\omega_{min}$ to high
frequency area above $\omega_{min}$. Roughly above 9000 \cm the
curves for different temperatures merge together, indicating that
the spectral weight is fully recovered. In the whole frequency range
the curve at 300 K is rather smooth, while subtle characters exist
in the spectra at 10 K. There appears a slight downturn centered at
700 \cm, manifesting the opening of the smaller CDW gap.
Additionally, another much striking suppression, verifying the
formation of the larger gap, occurs in a broad frequency range from
1500 \cm to 3500 \cm.

It should be noted that, in contrast to the first CDW transition
observed in the rare-earth based RTe$_3$ compounds, \cite{material}
which is pronounced to exist almost in the whole lanthanides series,
\cite{2CDW} the second CDW order is only reported in a fraction of
the RTe$_3$ materials. \cite{2CDW,opticalCeTe3} Apart from the prior
reports of the second CDW transition brought out by transport
\cite{2CDW} and ARPES measurement, \cite{ARPES2CDW} the present work
undoubtedly confirmed the coexistence of multiple CDW transitions in
ErTe$_3$, one of the heavy rare-earth tri-telluride. Our earlier
study on CeTe$_3$ \cite{opticalCeTe3} has provided an optical
evidence that multiple CDW orders also exist in the light rare-earth
tri-telluride. Therefore, it is natural to ask whether or not the
presence of the multiple CDW orders is a common feature in the
rare-earth based RTe$_3$ compounds. Then, more systematic
experimental investigations on various rare-earth based RTe$_3$
compounds would be necessary.

\section{\label{sec:level2}CONCLUSIONS}
To conclude, we have performed an optical investigation on the two
CDW phase transitions of ErTe$_3$. The well-defined plasma frequency
edge, which exhibits in the whole temperature scope, and the rather
high reflectivity in low frequency range both demonstrate the
well-behaved metallicity even in the CDW state. The depletion
features in R($\omega$) and corresponding peak characters in
$\sigma_1(\omega)$ both clearly reflect the development of the two
CDW orders upon cooling. Two energy gaps 2$\Delta_1$ = 343 meV and
2$\Delta_2$ = 110 meV at 10 K are obtained via the Drude-Lorentz fit
procedure. As the Drude component never disappears at low
temperatures, the FS in the ordered state is only partially gapped.
An optical estimation suggests that about half of the original FS is
gapped away. The broad suppression feature in R($\omega$) at room
temperature, which is yet above the first CDW transition, reflects
that fluctuation effect also plays an important role in this 2D
material.

\begin{center}
\small{\textbf{ACKNOWLEDGMENTS}}
\end{center}
This work was supported by the National Science Foundation of
China (10834013, 11074291), and the 973 project of the Ministry of
Science and Technology of China (2011CB921701).

\end{document}